\documentstyle[aps,prb,multicol,graphics]{revtex}
\begin{document}
\draft
\title{Screening, Coulomb pseudopotential, and superconductivity in  
alkali-doped Fullerenes}
\author{Erik Koch$^{(a,b)}$, Olle Gunnarsson$^{(a)}$, and
        Richard M.~Martin$^{(b)}$}
\address{${}^{(a)}$Max-Planck-Institut f\"ur Festk\"orperforschung,
         70506 Stuttgart, Germany}
\address{${}^{(b)}$Department of Physics, University of Illinois at
         Urbana-Champaign, Urbana, IL 61801, USA}
\date{\today}
\maketitle

\begin{abstract}
We study the static screening in a Hubbard-like model 
using quantum Monte Carlo. We find that the random phase 
approximation is surprisingly accurate almost up to 
the Mott transition. We argue that in alkali-doped Fullerenes
the Coulomb pseudopotential $\mu^\ast$ is not very much reduced 
by retardation effects. Therefore efficient screening is
 important in reducing $\mu^{\ast}$ sufficiently to allow 
for an electron-phonon driven superconductivity. In this 
way the Fullerides differ from the conventional picture, where 
retardation effects play a major role in reducing the 
electron-electron repulsion.
\end{abstract}
\pacs{74.70.Wz,71.10.Fd,71.20.Tx}

\begin{multicols}{2}
The random phase approximation (RPA) has been very widely used                
in solid state physics. It properly describes the screening when
the kinetic energy is much larger than the interaction 
energy.\cite{Pines} In the opposite limit, however, the RPA is 
qualitatively wrong. Little is known about the more interesting 
situation when the two energies are comparable.  In this 
paper we show for a Hubbard-like model that the  RPA gives a 
surprisingly accurate description of the static screening on 
the metallic side of a Mott transition until the system is 
close to the transition.

For conventional superconductors the electron-phonon interaction
leads to an effective electron-electron attraction. This 
interaction is counteracted by the strong Coulomb repulsion, 
which is, however, believed to be strongly reduced by retardation 
effects.\cite{retardation,screen} The resulting effective 
Coulomb interaction is described by the dimensionless Coulomb 
pseudopotential $\mu^{\ast}$, which is believed to be typically 
of the order 0.1. Here we argue that the situation for 
A$_3$C$_{60}$ (A= K, Rb) is different.  We find that retardation 
effects are rather inefficient. Therefore the screening of the 
Coulomb interaction becomes important for reducing the 
electron-electron repulsion. Thus, although the superconductivity 
in  A$_3$C$_{60}$ is driven by the electron-phonon 
interaction,\cite{Revmodphys} the origin of the strong 
reduction of $\mu^{\ast}$ is different from the current picture 
of conventional superconductors.  In the scenario we are 
putting forward, several puzzling phenomena find a natural 
explanation. In A$_3$C$_{60}$  (A= K, Rb) the transition
temperature $T_c$ is reduced by pressure.\cite{Fleming}  
For Cs$_3$C$_{60}$, however, which only under pressure becomes 
a superconductor, $T_c$ {\sl increases} with 
pressure.\cite{Palstra} This is consistent with the picture 
where $\mu^{\ast}$ is reduced by screening, since the 
screening is less efficient close to a Mott transition. 
Second, it was very early pointed out that the alkali phonons 
ought to couple efficiently to the electrons,\cite{Zhang} 
although later experiments showed that this was not the 
case.\cite{isotope} We show that efficient screening reduces 
the coupling to the alkali phonons.

We first discuss the screening in the RPA. In the random phase 
approximation it only costs kinetic energy to screen a test 
charge. In the limit where a typical Coulomb integral $U$ is 
large compared with the band width $W$, the kinetic energy 
cost of screening is relatively small compared with the 
potential energy gain, so the screening is efficient. This means 
that as a test charge $q$ is introduced on a site $c$, almost the 
same amount of electronic charge moves away from the site, 
leaving it almost neutral. This argument neglects, however, 
that when an electron leaves a site it has to find another 
site with a missing electron or there is a large Coulomb 
energy penalty.  Thus the RPA is accurate for small values of 
$U/W$, while it is {\sl qualitatively wrong} for large 
values. It is not clear what happens for intermediate values. 

To study the screening in  A$_3$C$_{60}$, we use a Hubbard-like 
model, including the three-fold degenerate $t_{1u}$ orbital: 
\begin{eqnarray}\label{eq:1}
H&&=\sum_{<ij>}\sum_{mm^{'}\sigma}t_{im,jm^{'}}
\psi^{\dagger}_{im\sigma^{\phantom{'}}}\psi^{\phantom{\dagger}}_{jm^{'}\sigma}\\
&&+\,U\sum_i\sum_{(m\sigma)<(m^{'}\sigma^{'})}n_{im^{\phantom{'}}\sigma}
n_{im^{'}\sigma'}
+q U\sum_{m\sigma} n_{c m \sigma} .\nonumber
\end{eqnarray}
The first term describes the kinetic energy, the second term 
the on-site Coulomb interaction and the third term the 
interaction with the test charge $q$ on site $c$. 
$\psi_{im,\sigma}$ annihilates an electron on site $i$ 
with orbital quantum number $m$ and spin $\sigma$, and 
$n_{im\sigma}=\psi^\dagger_{im\sigma} \psi^{\phantom
{\dagger}}_{im\sigma}$. The effect of orientational 
disorder\cite{TB,Mazin} is built into the hopping integrals 
$t_{im,jm^{'}}$. The band width is about $0.63\,eV$.
Multiplet effects are not included, but we remark that they 
tend to be counteracted by the Jahn-Teller effect 
which is also neglected. The test charge is assumed to interact 
with the electrons on the same site via the Coulomb integral $U$.
The system has three electrons per molecule, i.e.\ a 
half-filled $t_{1u}$ band.

We have investigated the model by using a lattice diffusion 
quantum Monte Carlo (QMC) method.\cite{QMC,deg} In this method 
a trial function $|\psi_T\rangle$ is constructed and allowed 
to diffuse towards the exact solution, under the constraint 
of a fixed node approximation.  $|\psi_T\rangle$ is obtained from 
a generalized Gutzwiller Ansatz\cite{Gutzwiller}
\begin{equation}\label{eq:2}
  |\Psi_T\rangle=g^D g_0^{n_c}|\Psi_0\rangle,
\end{equation}
where $|\Psi_0\rangle$ is a Slater determinant constructed 
from solutions of eqn.~(\ref{eq:1}) in the Hartree approximation, 
$D$ is the number of double occupancies in the system, and $n_c$ 
is the number of electrons on site $c$.  $g$ and $g_0$ are 
variational parameters. $g^D$ is the usual Gutzwiller factor 
while $g_0^{n_c}$ allows us to optimize the charge on the site 
with the test charge. In addition to the DMC calculation, we 
also perform a variational Monte Carlo calculation (VMC), and 
the energy is minimized as a function of $g$ and $g_0$.\cite{corrsmpl} 
In all cases the state is assumed to be paramagnetic. For 
$U/W\sim 2.5$ there is a transition to an antiferromagnetic 
Mott insulator,\cite{deg} where the screening is very inefficient. 
Here, however, we focus on $U/W<2.5$. We obtain the charge on 
site $c$ from the extrapolated estimator 
$n_c \approx 2n_c(DMC)-n_c(VMC)$, where $n_c(VMC)$ is the 
expectation value for the wave function (\ref{eq:2}) calculated 
by VMC and $n_c(DMC)$ is the mixed estimator from the DMC 
calculation. 

\begin{figure}
\centerline{
 \rotatebox{270}{\resizebox{!}{3in}{\includegraphics{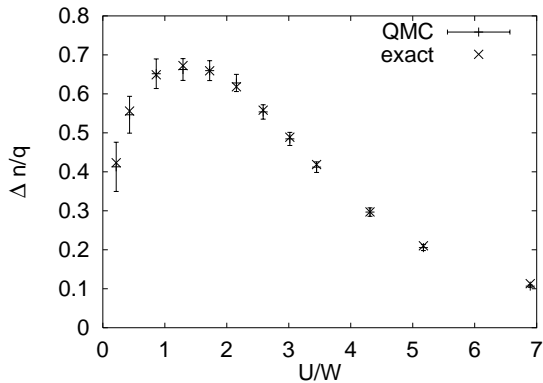}}}}
\caption[]{\label{fig1}   
 Screening charge $\Delta n$ on the site of the test charge 
($q=0.25\,e$) as a function of $U/W$, where $U$ is the Coulomb 
interaction and $W$ is the band width. Exact diagonalization 
and QMC calculations have been performed for four molecules 
(12 electrons). The figure shows that the QMC calculations 
are quite accurate over the whole range of $U/W$.}
\end{figure}

To test the accuracy of the approach, which involves the fixed-node 
approximation and uses the extrapolated estimator, we have 
compared the results of our QMC calculations with the the 
exact results from exact diagonalization of a system with 
four molecules (12 electrons). The comparison shown in 
Fig.~\ref{fig1} illustrates that the QMC calculations are 
quite accurate for the system we are analyzing here.

Since we are interested in the linear response, we should calculate
the effect of an infinitesimally small test charge $q$. Because 
of the statistical error in a QMC calculation it is, however, 
difficult to determine the response to a small perturbation. 
To get a good signal-to-noise ratio, we would therefore like 
to use as large a test charge as possible. To estimate how 
large we can make $q$ and still be in the linear response 
regime, we have performed Lanczos calculations for a range 
of different test charges. We find that for $q\leq 0.25\,e$ 
the response is practically linear. 

\begin{figure}[bt]
\centerline{
\rotatebox{270}{\resizebox{!}{3in}{\includegraphics{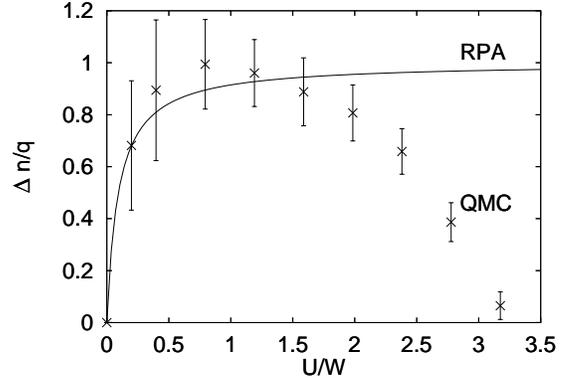}}}}
\caption[]{\label{fig2}   
 Screening charge $\Delta n$ on the site of the test charge 
($q=0.25\,e$) as a function of $U/W$, extrapolated to infinite 
cluster size. The full curve shows the screening charge in the 
RPA, obtained from Hartree calculations for the Hamiltonian 
(\ref{eq:1}). The crosses with errorbars give the results of 
the QMC calculations.  The RPA screening remains rather accurate 
up to $U/W\sim 2$, but fails badly for larger values of $U/W$. 
The screening is very efficient for $U/W\sim 0.5-2.0$.  }
\end{figure}

We have performed QMC calculations for larger clusters of 
$N_{\rm mol}$= 32, 48, 64, 72, and 108 molecules, where 
exact diagonalization is not possible. The screening charge 
$\Delta n_c=n_c(0)-n_c(q)$ was extrapolated to infinite cluster 
size, assuming a finite-size scaling of the form 
$\Delta n_c(N_{\rm mol})=\Delta n_c + \alpha/N_{\rm mol}$.
The results are shown in Fig.~\ref{fig2}.  For rather small 
values of $U/W$ ($\sim 0.5-1.0$), the RPA somewhat underestimates
the screening. Such a behavior is also found in the electron 
gas.\cite{Hedin} For intermediate values of $U/W$ ($\sim 1.0-2.0$) 
the RPA gives surprisingly accurate results. This is one of the main
results of this paper. For large $U/W$, the RPA rapidly 
becomes qualitatively wrong, as discussed earlier.  We are now 
in the position of addressing the superconductivity in A$_3$C$_{60}$.

In the theory of superconductivity, a dimensionless quantity 
$\mu^{\ast}$, the Coulomb pseudopotential, is introduced to 
describe the effects of the Coulomb repulsion. One introduces 
$\mu=UN(0)$, where $U$ is a typical screened Coulomb interaction 
and $N(0)$ is the density of states per spin at the Fermi energy. 
Retardation effects renormalize $\mu$ to $\mu^{\ast}$, and are 
described by ladder diagrams in the statically {\it screened} Coulomb 
interaction.\cite{retardation,screen} It is found that
\begin{equation}\label{eq:3}
 \mu^{\ast} =  {\mu\over 1+\mu{\rm ln}(\omega_{el}/\omega_{ph})}
          \sim {1\over {\rm ln}(\omega_{el}/\omega_{ph})},
\end{equation}
where $\omega_{el}$ and $\omega_{ph}$ are typical electron and phonon 
energy scales, respectively. Often $\mu$ln$(\omega_{el}/\omega_{ph})$ 
is substantially larger than unity.  In that limit the last 
part of eqn.~(\ref{eq:3}) holds, i.e.\ $\mu^\ast$ is
determined solely by retardation effects, independently of 
the screening, which only changes $\mu$.

In solid C$_{60}$ we have many narrow subbands ($\sim 0.5\,eV$ wide), 
spread over a range of about $30\,eV$. In the traditional 
approach one assumes that the relevant energy range extends 
over all this region. Summing the ladder diagrams in the
{\it screened} Coulomb interaction leads to a large renormalization 
of $\mu$. Exact results for a two-band model show, however,
that in the appropriate limits this approach greatly overestimates
the renormalization due to the upper sub band.\cite{mu} In the limit 
when a sub band is far away from the Fermi energy, the correct approach 
is to first project out the high energy degrees of freedom corresponding 
to  this sub band. This leads to an effective Hamiltonian, expressed
in terms of the {\sl unscreened} Coulomb matrix elements, which describes 
the low energy properties of the system. The main difference between the two 
approaches is the order in which high and low energy degrees 
of freedom are treated.  In the traditional approach the 
Coulomb interaction is screened first, which in particular 
involves the low energy degrees of freedom. After this the high
energy degrees of freedom are projected out. 
This approach involves uncontrolled approximations. Our
approach, instead, projects out the high energy degrees of freedom 
first, and it allows us to make statements about the importance of 
these degrees of freedom. Although these arguments were presented in 
the context of C$_{60}$, they are rather general. We now make more 
specific arguments for C$_{60}$ to provide further evidence that 
the retardation effects from higher sub bands are not very large.

From Auger measurements on K$_6$C$_{60}$, the Coulomb interaction 
$U$ between two holes in an otherwise full $t_{1u}$ band has 
been estimated to about 1.5 eV.\cite{U} This reduction of
$U$ from about 4 eV\cite{Antropov} for a free molecule to 
about $U_{\rm insul}=1.5$ eV for the the insulating solid, 
is mainly due to intramolecular processes and to polarization
of the molecules surrounding the two holes. Since the excitation 
energy of the relevant final state in the Auger experiment is rather 
small (about 1.5 eV), $U_{\rm insul}$ should contain the 
renormalization from all the higher sub bands, except possibly 
the ones closest to the $t_{1u}$ band. If we multiply  $U_{\rm insul}$ 
by $N(0)\sim 6$,\cite{susc} the result is a very large $\mu\sim 9$, 
much too large to allow for a phonon induced superconductivity 
unless $\mu$ is further reduced by other effects.

In K$_3$C$_{60}$ screening and retardation effects inside the 
$t_{1u}$ band become available. The argument against summing ladder
diagrams in the screened interaction were only justified for 
higher sub bands. Within the $t_{1u}$ band we
therefore rely on this conventional theory,\cite{retardation,screen} 
which in addition usually uses Thomas-Fermi or RPA screening. 
{\it A priori}, the use of RPA seems highly questionable for
these strongly correlated systems. Our calculations, however, 
support this approximatiom unless the system is close to a Mott 
transition.  Taking the long range Coulomb interaction into 
account, the RPA screening reduces $\mu$ to about $0.4$.\cite{mu} 
Including the additional retardation effects inside the $t_{1u}$ 
band according to eqn.~(\ref{eq:3}) finally renormalizes 
$\mu$ to $\mu^\ast \approx 0.3$. Thus the Coulomb pseudopotential 
is primarily reduced by screening and not by retardation effects. 
In contrast, using eqn.~(\ref{eq:3}) with $\omega_{el}\approx 15\,eV$ 
and $\omega_{ph}\approx 0.1\,eV$ would result in 
$\mu^\ast\lesssim 0.2$, practically {\sl independent of $\mu$}. 
A Coulomb pseudopotential $\mu^\ast\approx 0.3$ is substantially 
larger than for conventional superconductors,\cite{retardation} 
but it is not so large that it prevents the superconductivity 
from being driven by the electron-phonon interaction.\cite{c60m} 
Recent tunneling experiments give $\mu^\ast=0.329$ for 
Rb$_3$C$_{60}$.\cite{Dynes}

We now turn to the question how $T_c$ changes with the lattice 
constant $a$.  The main effect of increasing $a$ is to decrease 
the band-width $W$ and increase the density of states at the 
Fermi level $N(0)$. Using McMillan's formula, $T_c$ is given by 
\begin{equation}
  T_c={\omega_{ph}\over1.2}
 \exp\left[{-1.04(1+\lambda)\over\lambda-\mu^\ast(1+0.62\lambda)}
\right] ,
\end{equation}
with $\lambda=N(0)V$ the electron-phonon coupling constant. 
$\mu^\ast$ is calculated from $\mu=N(0)U_{\rm insul}\,(1-\gamma)$, 
where $U_{\rm insul}$ is a typical unscreened Coulomb matrix element 
and $\gamma=d\,n/d\,q$ describes the screening within the $t_{1u}$ band.  
Assuming that $\omega_{el}$ in Eq.\ (\ref{eq:3}) is large, $\mu^\ast$ is 
practically independent of the lattice constant $a$. Since $N(0)$ increases 
with decreasing $a$, the electron-phonon coupling $\lambda$ becomes 
stronger, increasing $T_c$.\cite{schluter} Assuming a small $\omega_{el}$, 
corresponding to the $t_{1u}$ band width, it is no longer true that 
$\mu^\ast$ is independent of $\mu$. However, if the RPA is valid, 
$\mu$ is almost independent of the lattice constant, since the 
increase in $N(0)$ is counteracted by a slightly more efficient 
screening $\gamma_{RPA}$ (cf.~Fig.~\ref{fig2}). Hence also in this 
scenario we find that $T_c$ increases with $a$. But what happens 
when the lattice constant $a$ becomes large enough that we enter 
the region where the screening starts to break down? Then $\mu$ 
will start to increase considerably with $a$. Assuming a large $\omega_{el}$, 
$\mu^\ast$ is still independent of $\mu$, and therefore $T_c$ should 
keep increasing. For small $\omega_{el}$, on the other hand, $\mu^\ast$ will 
start to rapidly increase with $a$, leading to a steep drop in $T_c$. 
This resembles the anomalous behavior observed in Cs$_3$C$_{60}$: 
it only becomes superconducting under pressure, with $T_c$ rapidly 
decreasing with increasing lattice constant.\cite{Palstra} 

It might appear that efficient screening is not really helpful for 
superconductivity. Phonons couple to the electrons by perturbing the potential 
seen by the electrons.\cite{retardation,screen} An example being the 
longitudinal modes of a jellium. Efficient screening tends to weaken the 
coupling to such phonons, since it reduces the perturbation considerably. To 
some extent, such a reduction also seems to be at work in C$_{60}$. 
Initially it was expected that the coupling to the alkali phonons would
be very strong.\cite{Zhang} Each C$_{60}$ molecule is surrounded by 14 
alkali ions with relatively weak force constants.  When an electron 
arrives on a C$_{60}$ molecule one would therefore expect that the 
surrounding alkali ions respond strongly. This was, however, not 
confirmed by experiment. For instance, an alkali isotope effect could 
not be observed within the experimental accuracy.\cite{isotope} This 
finding can be naturally understood as an effect of the efficient 
screening found in our calculations. When an electron arrives on a 
C$_{60}$ molecule, other electrons leave the molecule, which thus 
stays almost neutral. The alkali ions then only see a small change in 
the net charge and therefore couple weakly.  In a similar way it follows 
that intramolecular phonons of A$_g$ symmetry couple weakly. An A$_g$ 
phonon shifts all the $t_{1u}$ levels on a given molecule in the same 
direction. This shift of the center of gravity can be screened very 
efficiently by transferring charge from the molecules where the levels 
move upwards to those where they move downwards. The modes that are 
important for the superconductivity in solid C$_{60}$ are, however, different.
An intramolecular H$_g$ phonon does {\em not} shift the center of gravity
of the $t_{1u}$ level. Thus the H$_g$ phonons are {\em not} screened by 
the transfer of charge. Hence for these phonons the efficient screening
serves to reduce $\mu^\ast$ without affecting the electron-phonon coupling.

To summarize, we have calculated the static screening of a point charge for a 
Hubbard-like model using quantum Monte Carlo. We find that the RPA is 
surprisingly accurate up to values of $U/W$ fairly close to the Mott transition.
For larger $U/W$ the screening rapidly breaks down. This result should have
quite general implications for the physics of systems close to a Mott 
transition. Here we have studied the consequences for the superconductivity
in the alkali-doped Fullerenes. We have provided arguments that for 
A$_3$C$_{60}$ (A= K, Rb) retardation effects are very inefficient in reducing 
the electron-electron repulsion. Instead, and unlike for textbook 
superconductors, screening is mainly responsible for the reduction of the 
Coulomb pseudopotential $\mu^\ast$. This results in a $\mu^{\ast}$ small 
enough that the electron-phonon interaction can drive the superconductivity. 
Nevertheless $\mu^{\ast}$ is substantially larger than for conventional 
superconductors, in agreement with recent experiments.
This scenario is quite different from the conventional picture of a 
superconductor, where the retardation effects are believed to play the central 
role in reducing $\mu^{\ast}$. It explains quite naturally the anomalous 
pressure dependence of $T_c$ found for Cs$_3$C$_{60}$ and the absence of a 
strong coupling to the alkali phonons. It also predicts that the coupling to 
the A$_g$ phonons is strongly reduced by screening effects. 
Finally, our results let us understand the surprising fact that $T_c$ peaks 
for systems close to the Mott transition, where the density of states is 
large, but the screening has not yet started to become inefficient.

This work has been supported by the Alexander-von-Humboldt-Stiftung under the
Feodor-Lynen-Program and the Max-Planck-Forschungspreis, and by the Department
of Energy, grant DEFG 02-96ER45439.

\end{multicols}
\end{document}